\documentclass[useAMS,usenatbib]{mn2e}
\usepackage[dvips]{epsfig}
\usepackage[]{natbib}
\usepackage[]{fixltx2e}
\title[Gas inflows towards the nucleus of NGC\,1667]{Gas inflows towards the nucleus of the Seyfert 2 galaxy NGC\,1667}
\author[A. Schnorr-M\"uller et al.]
  {Allan Schnorr-M\"uller,$^1$ Thaisa Storchi-Bergmann,$^1$ Fabricio Ferrari$^2$, Neil M. Nagar,$^3$\\
  $^1$Instituto de F\'isica, Universidade Federal do Rio Grande do Sul, 91501-970, Porto Alegre, RS, Brazil\\
  $^2$Instituto de Matem\'atica, Estat\'istica e F\'isica, Universidade Federal do Rio Grande (FURG), 96201-900, Rio Grande, RS, Brazil\\
  $^3$Astronomy Department, Universidad de Concepci\'on, Casilla 160-C, Concepci\'on, Chile\\}

\pagerange{\pageref{firstpage}--\pageref{lastpage}} \pubyear{2012}

\begin{document}

\label{firstpage}
\maketitle

\begin{abstract}

We use optical spectra from the inner 2\,$\times$\,3\,kpc$^2$ of the Seyfert\,2 galaxy NGC\,1667, obtained with the GMOS integral field spectrograph on the Gemini South telescope at a spatial resolution of $\approx$\,240\,pc, to assess the feeding and feedback processes in this nearby AGN. We have identified two gaseous kinematical components in the emission line profiles: a broader component ($\sigma$\,$\approx$\,400\,km\,s$^{-1}$) which is observed in the inner 1--2\arcsec\ and a narrower component ($\sigma$\,$\approx$\,200\,km\,s$^{-1}$) which is present over the entire field-of-view. We identify the broader component as due to an unresolved nuclear outflow. The narrower component velocity field shows strong isovelocity twists relative to a rotation pattern, implying the presence of strong non-circular motions. The subtraction of a rotational model reveals that these twists are caused by outflowing gas in the inner $\approx$\,1\arcsec, and by inflows associated with two spiral arms at larger radii. We calculate an ionized gas mass outflow rate of $\dot{M}_{out}$\,$\approx$\,0.16\,M$_{\odot}$\,yr$^{-1}$. We calculate the net gas mass flow rate across a series of concentric rings, obtaining a maximum mass inflow rate in ionized gas of $\approx$\,2.8\,M$_{\odot}$\,year$^{-1}$ at 800\,pc from the nucleus, which is two orders of magnitude larger than the accretion rate necessary to power this AGN. However, as the mass inflow rate decreases at smaller radii, most of the gas probably will not reach the AGN, but accumulate in the inner few hundred parsecs. This will create a reservoir of gas that can trigger the formation of new stars.  

\end{abstract}


\section{Introduction}

It is widely accepted that the radiation emitted by an active galactic nucleus (AGN) is a result of accretion onto the central supermassive black hole (hereafter SMBH). However, understanding how mass is transferred from kiloparsec scales down to nuclear scales of tens of parsecs and the mechanisms involved has been a long-standing problem in the study of nuclear activity in galaxies.  Observational studies suggest that nuclear dust structures such as spirals, filaments and disks are tracers of the feeding channels to the AGN, as a strong correlation between nuclear activity and the presence of nuclear dust structures have been found in early type galaxies \citep{lopes07}. Simulations showed that, if a central SMBH is present, spiral shocks can extend all the way to the SMBH vicinity and generate gas inflow consistent with the observed accretion rates \citep{maciejewski04a,maciejewski04b}. 

In order to assess if nuclear spirals are capable of channeling gas inwards to feed the SMBH, our group has been mapping gas flows in the inner kiloparsec of nearby AGN using optical and near-infrared integral field spectroscopic observations. So far, we have observed gas inflows along nuclear spirals in NGC\,1097 \citep{fathi06}, NGC\,6951 \citep{thaisa07}, NGC\,4051 \citep{rogemar08}, M\,79 \citep{rogemar13}, NGC\,2110, \citep{allan14a} and NGC\,7213 \citep{allan14b}. We also observed gas inflows in the galaxy M\,81 \citep{allan11}, where the inflow was mostly traced by dust lanes and in NGC\,3081 \citep{allan16}, where a nuclear bar is feeding the AGN. Gas inflows have also been observed by other groups. Near-infrared integral field spectroscopic observations revealed inflows along nuclear spiral arms in NGC\,1097 \citep{davies09}, NGC\,5643 \citep{davies14} and NGC\,7743 \citep{davies14}, and gas inflow along a bar in NGC\,3227 \citep{davies14}. Recent ALMA observations releaved streaming motions along nuclear spirals in NGC\,1433 \citep{combes13} and NGC\,1566 \citep{combes14}. Gas inflows have been observed in CO in NGC\,2782 \citep{hunt08}, NGC\,3147 \citep{casasola08}, NGC\,3627 \citep{casasola11}, NGC\,4579 \citep{burillo09} and NGC\,6574 \citep{krieg08}.

In the present work, we report results obtained from optical integral field spectroscopic observations of the nuclear region of NGC\,1667, a spiral galaxy with Hubble type SAB(r)c which harbors a Seyfert 2 AGN enclosed by H\,II regions \citep{delgado97}. NGC\,1667 is located at a distance of 61.0\,Mpc (NED\footnote{NASA/IPAC extragalactic database}), corresponding to a scale of 296\,pc/\arcsec. 

The present paper is organized as follows. In Section \ref{Observations} we describe the observations and reductions. In Section \ref{Results} we present the procedures used for the analysis of the data and the subsequent results. In section \ref{Discussion} we discuss our results and present estimates of the mass inflow rate from two distinct methods and in Section \ref{Conclusion} we present our conclusions.

\begin{figure*}
\includegraphics[scale=0.8]{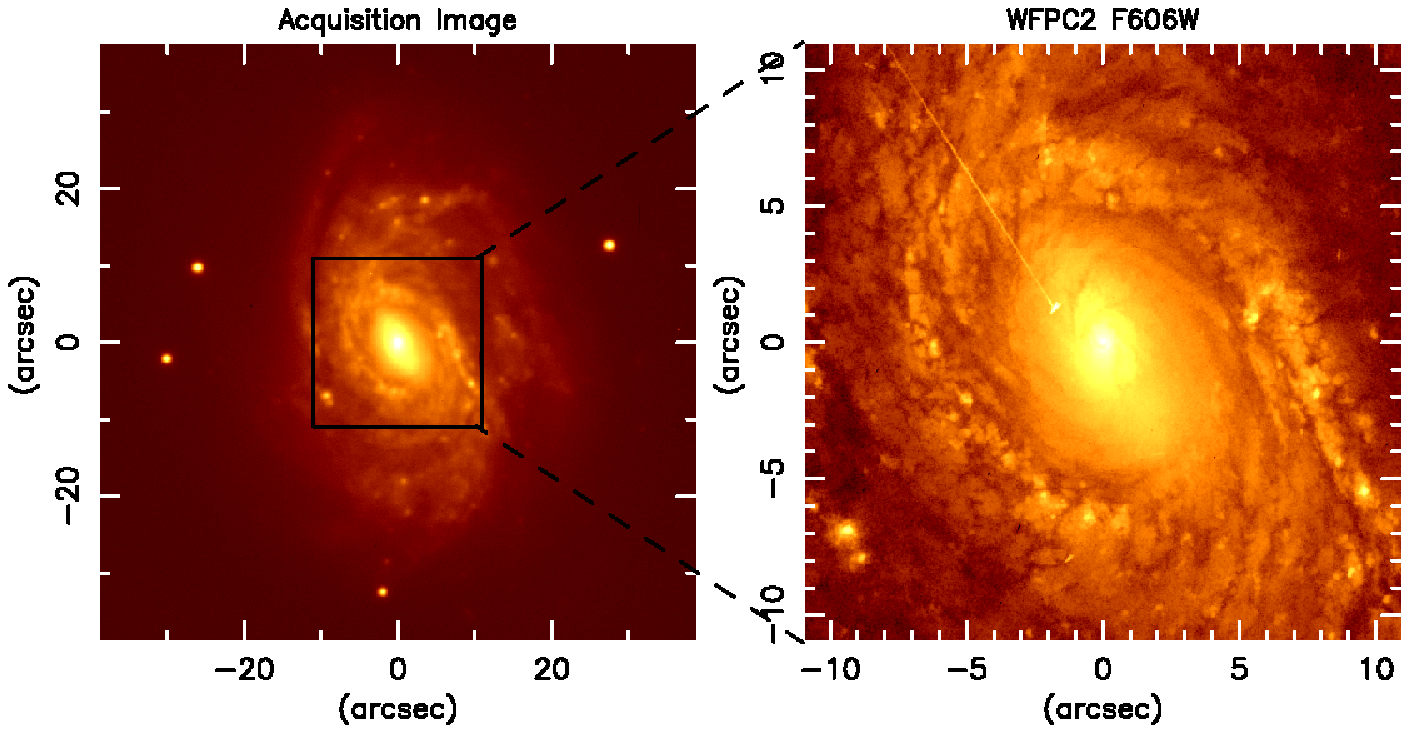}
\includegraphics[scale=0.8]{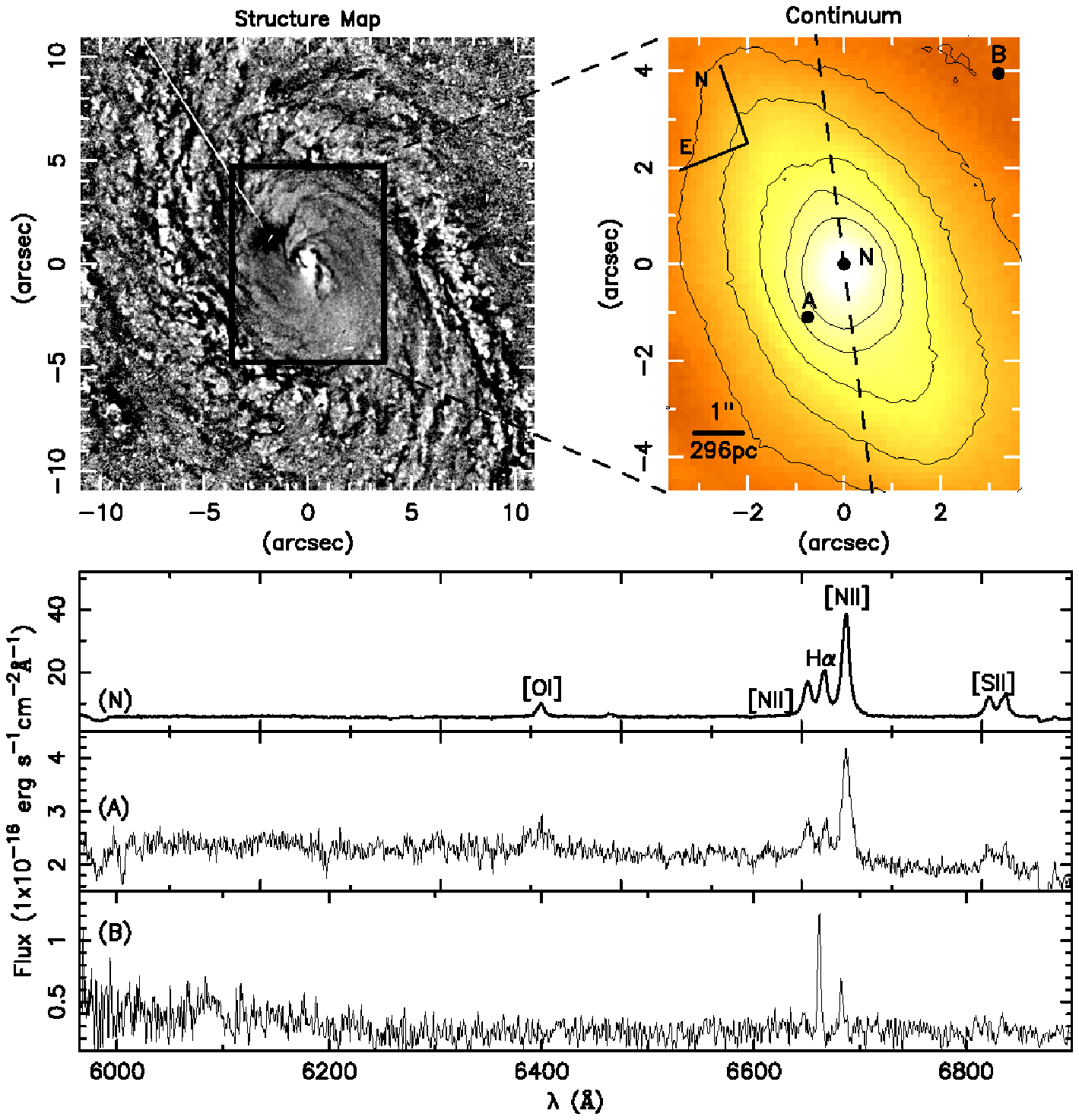}
\caption[Large scale image of NGC1667]{Top left: acquisition image of NGC1667. Top right: WFPC2 image. Middle left: structure map. The rectangle shows the field of the IFU observation. Middle right: continuum image from the IFU spectra (flux in erg\,cm$^2$\,s$^{-1}$ per pixel). The dashed black line indicates the position angle of the photometric major axis of the galaxy (PA\,=\,165\ensuremath{^\circ}), from the 2MASS survey \citealt{twomass}). Bottom: spectra corresponding to the regions marked as N, A and B in the IFU image. All images have the same orientation. The thin line in the HST image and structure map is an artifact.}
\label{fig1}
\end{figure*}

\section {Observations and Reductions}\label{Observations}

The observations were obtained with the Integral Field Unit of the Gemini Multi Object Spectrograph (GMOS-IFU) at the Gemini South telescope on the night of January 26, 2011 (Gemini project GS-2010B-Q-19). The observations consisted of two adjacent IFU fields (covering 7\,$\times$\,5\,arcsec$^{2}$ each) resulting in a total angular coverage of 7\,$\times$\,10\,arcsec$^{2}$ around the nucleus. Six exposures of 350 seconds were obtained for each field, slightly shifted and dithered in order to correct for detector defects after combination of the frames. The seeing during the observation was 0\farcs8, as measured from the FWHM of a spatial profile of the calibration standard star. This corresponds to a spatial resolution at the galaxy of 237\,pc. The photometric major axis is oriented along the position angle (PA) 165\ensuremath{^\circ} (\citealt{twomass}, derived from a Near-IR image of the galaxy from the Two Micron All-Sky Survey (2MASS)).

The selected wavelength range was 5600-7000\,\r{A}, in order to cover the H$\alpha$+[N\,II]\,$\lambda\lambda$6548,6583\,\r{A} and [S\,II]\,$\lambda\lambda$6716,6731\,\r{A} emission lines. The observations were obtained with the grating GMOS R400-G5325, set to central wavelength of either $\lambda$\,6500\,\r{A} or $\lambda$\,6550\,\r{A}, at a resolving power of R\,$\approx$\,2000.

The data reduction was performed using specific tasks developed for GMOS data in the \textsc{gemini.gmos} package as well as generic tasks in \textsc{iraf}\footnote{\textit{IRAF} is distributed by the National Optical Astronomy Observatories, which are operated by the Association of Universities for Research in Astronomy, Inc., under cooperative agreement with the National Science Foundation.}. The reduction process comprised bias subtraction, flat-fielding, trimming, wavelength calibration, sky subtraction, relative flux calibration, building of the data cubes at a sampling of 0.1\arcsec$\,\times\,$0.1\arcsec, and finally the alignment and combination of 12 data cubes.

\section{Results}\label{Results}

In Fig.\,\ref{fig1} we present in the upper left panel the acquisition image of NGC\,1667 and in the upper right panel an image of the inner 22\arcsec\,$\times\,$22\arcsec\ of the galaxy obtained with the Wide Field Planetary Camera 2 (WFPC2) through the filter F606W aboard the Hubble Space Telescope (HST). In the middle left panel we present a structure map of the WFPC2 HST image (see \citealt{lopes07}). The rectangle in the middle left panel shows the field-of-view (hereafter FOV) covered by the IFU observations. Nuclear spiral arms traced by (dark) dust lanes are a prominent feature observed inside this rectangle. In the middle right panel we present an image from our IFU observations obtained by integrating the continuum flux within a spectral window from $\lambda$6470\,\r{A} to $\lambda$6580\,\r{A}. The isophotes in this image are extended roughly along the North-South direction, tracing the position of a nuclear bar \citep{laine02}. In the lower panel we present three spectra of the galaxy corresponding to locations marked as A, B and N in the IFU image, selected to be representative of the observed spectra.  These spectra were extracted within apertures of 0\farcs3\,$\times\,$0\farcs3. 

The spectrum corresponding to the nucleus (marked as N in Fig.\,\ref{fig1}) is typical of the inner 1\arcsec, showing strong narrow [O\,I]\,$\lambda$$\lambda$\,6300,6363\,\r{A}, [N\,II]\,$\lambda$$\lambda$6548,6583\,\r{A}, H$\alpha$ and [S\,II]\,$\lambda$$\lambda$6717,6731\,\r{A} emission lines. In this spectrum the lines have a broad base and a narrow core, a shape distinct from a single Gaussian profile. The spectrum from location A is similar to the nuclear spectrum, and typical of the spectra of regions between 1\arcsec\ to 3\arcsec\ from the nucleus. The spectrum from location B is typical from the borders of the FOV ($\approx$3\farcs5 from the nucleus), where the [N\,II] and H$\alpha$ lines are very narrow and the [N\,II]/H$\alpha$ line ratio is typical of an HII region.

\subsection{Measurements}

\begin{figure*}
\includegraphics[scale=0.7]{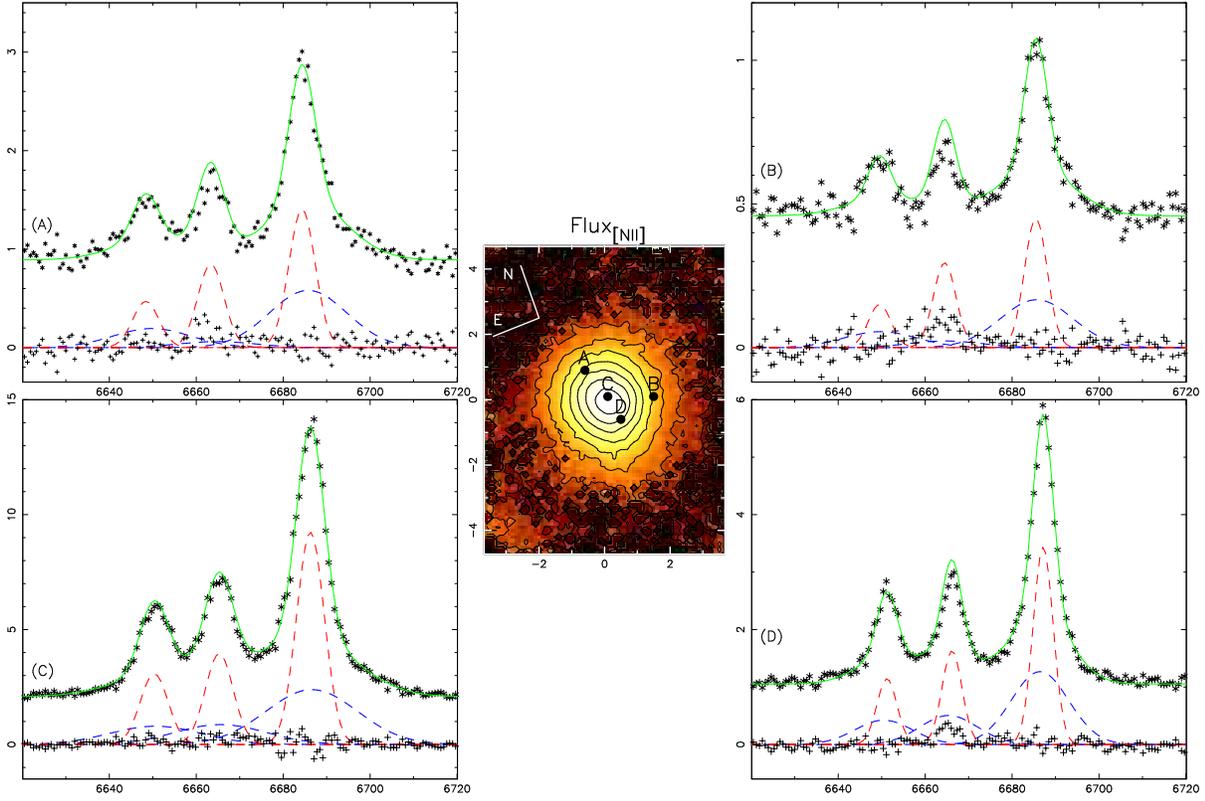}
\caption{Two Gaussian fits to the [N\,II] and H$\alpha$ emission lines from four different regions, labeled as A, B, C and D. The asterisks correspond to data points, the solid lines to the fit, crosses to the residuals and the dashed lines to the narrower and broader components. }
\label{figgauss}
\end{figure*}

The gaseous centroid velocities, velocity dispersions and the emission-line fluxes were obtained through the fit of Gaussians to the [N\,II], H$\alpha$, [O\,I] and [S\,II] emission lines. In order to reduce the number of free parameters when fitting the [N\,II] and H$\alpha$ lines, we adopted the following physically motivated constraints: 
\begin{enumerate}
\item Flux$_{[N\,II]\,\lambda6583}$/Flux$_{[N\,II]\,\lambda6548}=2.98$, in accordance with the ratio of
their transition probabilities \citep{osterbrock06};
\item The H$\alpha$, [N\,II]\,$\lambda$6583 and [N\,II]\,$\lambda$6548 lines have the same centroid velocity and FWHM;
\end{enumerate}

Additionaly, we performed an adaptive spatial binning of our datacube using the Voronoi binning technique \citep{voronoi} so that the [N\,II]\,$\lambda$6583\r{A} line has a signal-to-noise ratio of at least 3 in each pixel. This allowed us to measure this line in the entire FOV without losing any significant spatial information as binning was only necessary near the borders of the FOV and the number of pixels in each bin is small. We did not perform an adaptive spatial binning before measuring the [S\,II] lines as there is no signal in these lines outside the inner 2\arcsec. Uncertainties in the kinematic parameters were estimated using Monte Carlo simulations based on the best-fitting absorption spectra.

A visual inspection of the fits to the spectra showed that in the inner $\approx$\,2\arcsec\ the [N\,II], H$\alpha$ and [S\,II] emission line profiles have a broad base and a narrow core, a shape which could not be reproduced by a single Gaussian profile. Therefore, in addition to performing a single Gaussian fit to the emission line profiles in the entire datacube, we also performed a two Gaussian fit to the line profiles in the inner 2\arcsec. The two component fit resulted in a component tracing the narrow core of the profile, with velocity dispersion between 100-180\,km\,s$^{-1}$ and centroid velocities roughly the same as those resulting from the single Gaussian fit, and a second component tracing the broad base, with velocity dispersion between 300-600\,km\,s$^{-1}$ and centroid velocities close to systemic. We will refer to these components as the ``narrower component" and the ``broader component" from here on. Fits where one of the components contributed less than 10\% to the total flux of each emission line where discarded. In Fig.\,\ref{figgauss} we show four examples of the two Gaussians fits to the [N\,II] and H$\alpha$ lines. We followed the same procedure when fitting the [S\,II] lines, but constraining the centroid velocity of each component as equal to the value determined from the two Gaussian fit to the [N\,II] lines. In the next sections we present and discuss the results from both the Single Gaussian and Two Gaussians fits.       

In order to measure the stellar kinematics, we employed the Penalized Pixel Fitting technique (pPXF, \citealt{cappellari04}) and using the \citet{bruzual03} stellar population models as templates to fit the stellar continuum from 5700\,\r{A} to 6600\,\r{A}, where numeroues absorption features are present, the strongest being the Na\,I doublet at $\lambda\lambda$5890/5896\,\r{A}. Using the Voronoi Binning technique \citep{voronoi} we binned the datacube to achieve a signal-to-noise ratio of at least 5 in the continuum near the Na\,I doublet. The uncertainties in the kinematic parameters were estimated using Monte Carlo simulations based on the best-fitting absorption spectra.

\subsection{Uncertainties}\label{uncertainties}

\begin{figure*}
\includegraphics[scale=1.1]{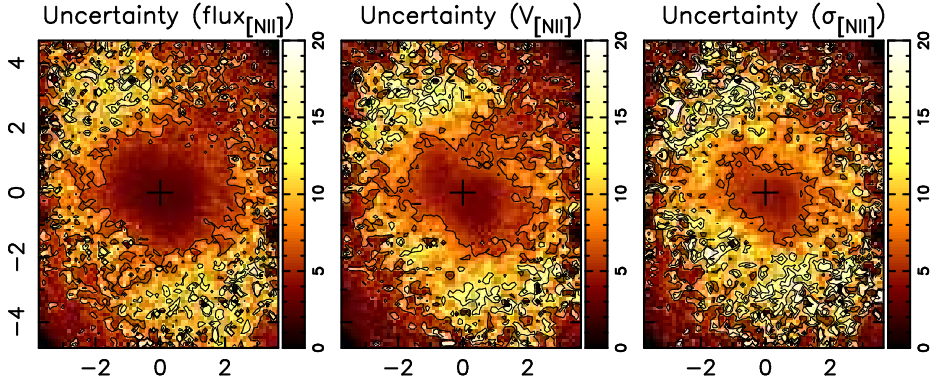}
\caption[Gaseous Kinematics]{Uncertainties in flux (\%), centroid velocity (km\,s$^{-1}$) and velocity dispersion (km\,s$^{-1}$) for the [N\,II] emission line.}
\label{figerror}
\end{figure*}

To test the robustness of the fits and estimate the uncertainties in the quantities measured from each spectrum in our datacube, we performed Monte Carlo simulations in which Gaussian noise was added to the observed spectrum. For each spaxel, the noise added in each Monte Carlo iteration was randomly drawn from a Gaussian distribution whose standard deviation matches that expected from the Poissonian noise of that spaxel. One hundred iterations were performed and the estimated uncertainty in each parameter - line center, line width, and total flux in the line - was derived from the $\sigma$ of the parameter distributions yielded by the iterations. In Fig.\,\ref{figerror} we show the uncertainties in the measurement of the [N\,II] emission lines. We note that the uncertainties decrease near the edges of the FOV because we observe strong emission from H\,II regions in the nuclear ring there (see spectrum B in Fig\,\ref{fig1}).  Uncertainties in the fluxes, velocities, and velocity dispersions of the H$\alpha$ and [S\,II] lines are similar to those of the [N\,II] line. Uncertainties in the gas density are of the order of 35$\%$. Uncertainties in the stellar velocity and velocity dispersion are of the order of 20\,km\,s$^{-1}$ and 25\,km\,s$^{-1}$ respectively.

\subsection{Stellar kinematics}

\begin{figure}
\includegraphics[scale=1.1]{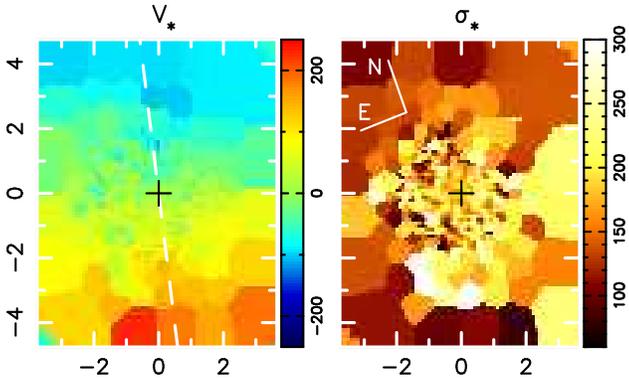}
\caption[Gaseous Kinematics]{Stellar centroid velocity (km\,s$^{-1}$) and velocity dispersion (km\,s$^{-1}$).}
\label{figstars}
\end{figure}

The stellar velocity field (left panel of Fig.\,\ref{figstars}) displays a rotation pattern in which the N side of the galaxy is approaching and the S side is receding. Under the assumption that the spiral arms are trailing, it can be concluded that the near side of the galaxy is to the E, and the far side is to the W. A systemic velocity of 4650\,km\,s$^{-1}$ (see section.\,\ref{discuss-stars} for details on how this value was determined) was subtracted from the centroid velocity maps. The stellar velocity dispersion (right panel of Fig.\,\ref{figstars}) varies between 100\,km\,s$^{-1}$ and 400\,km\,s$^{-1}$. 

\subsection{Gaseous Kinematics}\label{kinematics}

\subsubsection{Single Gaussian fit}

\begin{figure*}
\includegraphics[scale=1.2]{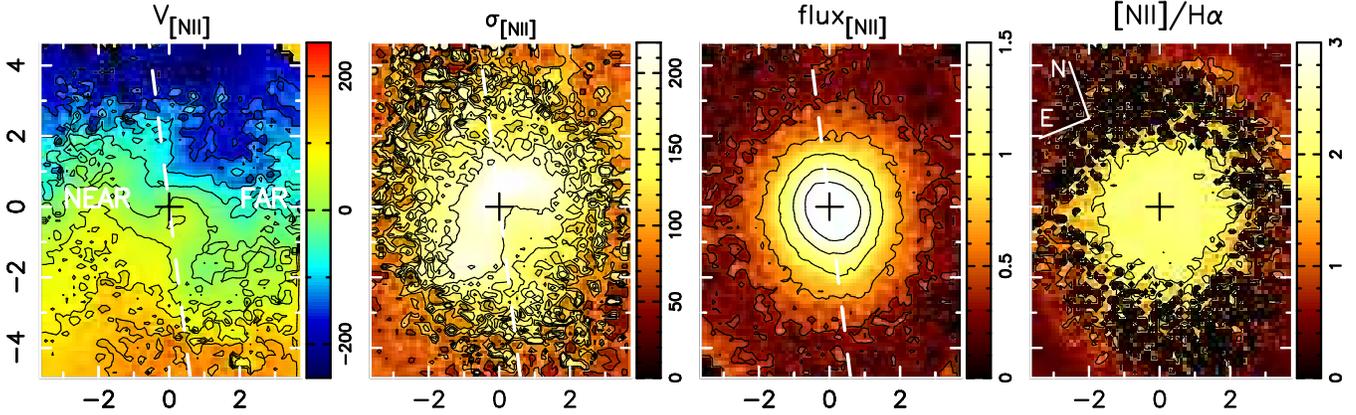}
\caption[Gaseous Kinematics]{Single Gaussian fit: gaseous centroid velocities (km\,s$^{-1}$),  velocity dispersion, logarithm of the [N\,II] emission line flux (in units of 10$^{-17}$\,erg\,cm$^2$\,s$^{-1}$ per spaxel) and [N\,II]/H$\alpha$ ratio.}
\label{figsingle}
\end{figure*}

In Fig.\,\ref{figsingle} we show centroid velocity (km\,s$^{-1}$),  velocity dispersion, flux distribution, and [N\,II]/H$\alpha$ ratio maps obtained from our single Gaussian fit to the [N\,II] and H$\alpha$ emission lines. A systemic velocity of 4650\,km\,s$^{-1}$ was subtracted from the centroid velocity maps. The gaseous velocity field is dominated by non-circular motions; note specially the distortions in the isovelocity curves in the inner 2\arcsec, especially close to the minor axis. Asymmetries are also visible in the centroid velocity field, for example a steep increase in velocity observed to the northwest, where the centroid velocity rises from $\approx$\,150\,km\,s$^{-1}$ at 2\farcs5 to $\approx$\,200\,km\,s$^{-1}$ at 4\arcsec\ while no similar increase is observed to the south. 

The velocity dispersion maps show low values, between 60-100\,km\,s$^{-1}$, near the edges of the FOV, with an increase to 130-220\,km\,s$^{-1}$ in the inner 2\farcs5. The largest values ($\approx$\,200\,km\,s$^{-1}$) are observed in a ``C'' shaped region, extending from the southeast to the northwest of the nucleus.

\subsubsection{Two Gaussians fit}

\begin{figure*}
\includegraphics[scale=1.2]{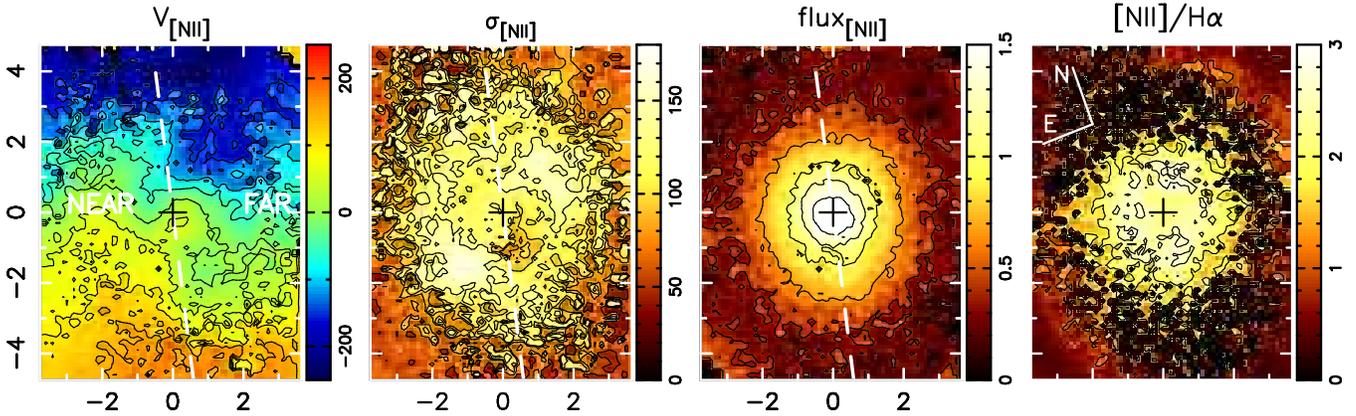}
\caption[Gaseous Kinematics]{Narrower component: gaseous centroid velocities (km\,s$^{-1}$),  velocity dispersion, logarithm of the [N\,II] emission line flux (in units of 10$^{-17}$\,erg\,cm$^2$\,s$^{-1}$ per spaxel), and [N\,II]/H$\alpha$ ratio. This map was constructed combining the narrower component and the single Gaussian maps (where only one component was fitted).}
\label{fignarrow}
\end{figure*}

\begin{figure*}
\includegraphics[scale=1.2]{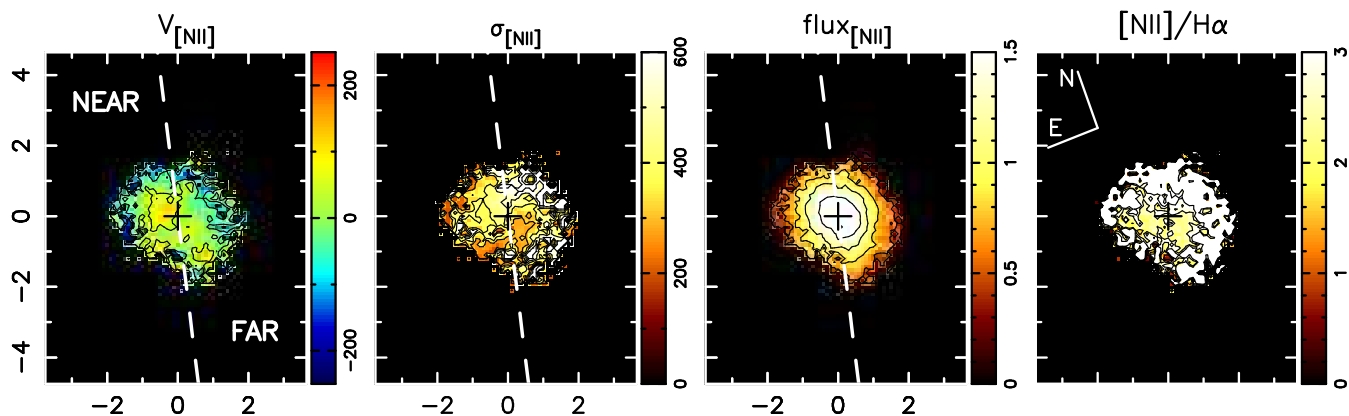}
\caption[Gaseous Kinematics]{Broader component: gaseous centroid velocities (km\,s$^{-1}$),  velocity dispersion, logarithm of the [N\,II] emission line flux (in units of 10$^{-17}$\,erg\,cm$^2$\,s$^{-1}$ per spaxel), and [N\,II]/H$\alpha$ ratio.}
\label{figbroad}
\end{figure*}

The narrower component (Fig.\,\ref{fignarrow}) velocity field is essentially the same as the one obtained from the single Gaussian fit, the differences between the two  do not exceed 20\,km\,s$^{-1}$. The narrower component centroid velocity map shows the same features as the single Gaussian map, although the velocity dispersions are on average 30\,km\,s$^{-1}$ lower and the ``C'' shaped region extends further to the northwest and southeast. The broader component centroid velocity map (Fig.\,\ref{figbroad}) shows velocities close to systemic, except for a small region to the northeast of the nucleus, where velocities of $\approx$\,100\,km\,s$^{-1}$ are observed. The velocity dispersion ranges from 200\,km\,s$^{-1}$ to 600\,km\,s$^{-1}$, the highest velocity dispersion being observed northwest of the nucleus and the lowest south of the nucleus.

\subsection{Line fluxes and excitation of the emitting gas}\label{fluxgas}

\begin{figure*}
\begin{center}
\includegraphics[scale=1.2]{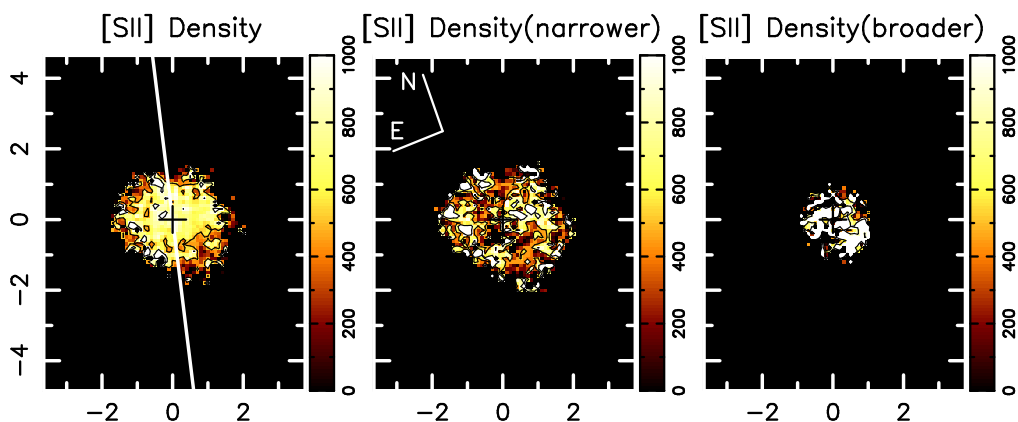}
\caption[Line ratio maps]{Gas density (cm$^{-3}$) obtained from the single component fit and gas density of the narrower and broader components.}
\label{figraz}
\end{center}
\end{figure*}

We show the [N\,II] flux distribution for the single Gaussian fit in the center right panel of Fig.\,\ref{figsingle}. The [N\,II] flux distribution for the narrower and broader components are shown in the center right panels of Fig.\,\ref{fignarrow} and Fig.\,\ref{figbroad} respectively. The Single Gaussian and narrower component flux and excitation maps present essentially the same structures, thus we only discuss the narrower component maps from here on. 

The narrower component [N\,II] flux distribution shows an elongation along the major axis, a decrease of the emission in a ring at $\approx$\,2\arcsec\ from the nucleus and an increase near the borders of the FOV. The [N\,II]/H$\alpha$ map of the narrower component presents values of the order of 0.5 near the edges of the FOV. This is consistent with photoionization by young stars, which is expected as these regions correspond to the inner part of a star forming ring (see Fig.\,\ref{fig1}). The ratio increases to 1.5-2.5 in the inner 2\arcsec, values consistent with photoionization by a Seyfert nucleus. 

The broader component [N\,II] flux distribution shows an elongation towards the northeast. The broader component [N\,II]/H$\alpha$ ratio varies between 2-3, higher than the narrower component, which suggests a larger contribution of excitation by shocks in the gas in this component.     

In Fig.\ref{figraz} we present the gas density map for the single Gaussian fit, narrower component and broader component, obtained from the [SII]\,$\lambda\lambda$6717/6731\,\r{A} line ratio assuming an electronic temperature of 10000K \citep{osterbrock06}. The narrower component gas density varies between 300-800\,cm$^{-3}$. The broader component density reaches $\approx$\,1000\,cm$^{-3}$.

\section{discussion}\label{Discussion}

\subsection{The stellar kinematics}\label{discuss-stars}

\begin{figure*}
\includegraphics[scale=1.2]{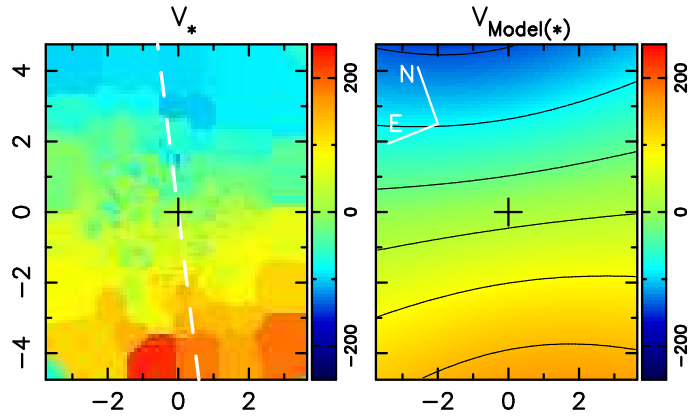}
\caption[Stellar Kinematics]{Stellar velocity field (km\,s$^{-1}$) and modeled velocity field (km\,s$^{-1}$). The dashed white line indicates the position angle of the major axis of the galaxy.}
\label{figmodel}
\end{figure*}

In order to obtain the value of the systemic velocity and the rotation velocity field of the disk of the galaxy, we modeled the stellar velocity field assuming a spherical potential with pure circular motions, with the observed radial velocity at a position ($R,\psi$) in the plane of the sky given by \citep{bertola91}:
\footnotesize
\begin{displaymath}
V=V_{s}+\frac{ARcos(\psi-\psi_{0})sin(\theta)cos^{p}\theta}{\{R^{2}[sin^{2}(\psi-\psi_{0})+cos^{2}\theta cos^{2}(\psi-\psi_{0})]+c^{2}cos^{2}\theta \}^{p/2}} 
\end{displaymath}
\normalsize
where $\theta$ is the inclination of the disk (with $\theta$\,=\,0 for a face-on disk), $\psi_{0}$ is the position angle of the line of nodes, $V_{s}$ is the systemic velocity, $R$ is the radius, $A$ is the amplitude of the rotation curve (at large radii), $c$ is a concentration parameter regulating the compactness of the region with a strong velocity gradient and $p$ regulates the inclination of the flat portion of the velocity curve (at the largest radii). We assumed the kinematical center to be cospatial with the peak of the continuum emission. We adopted an inclination of i\,=\,48\ensuremath{^\circ} \citep{radovich96}. A Levenberg-Marquardt least-squares minimisation was performed to determine the best fitting 
parameters.

The resulting parameters $A$, $c$, and $p$ are $380\,\pm$16\,km\,s$^{-1}$, $7\farcs3\pm0.4$ and $1.0\,\pm0.1$ respectively. The position angle of the line of nodes was fixed as 165\ensuremath{^\circ}$\pm1$, which is the position angle of the large scale photometric major axis \citep{twomass}. This was done because a large range of position angles resulted in good fits due to the limited number of data points. The systemic velocity corrected to the heliocentric reference frame is $4570\,\pm$15\,km\,s$^{-1}$ (taking into account both errors in the measurement and the fit). This value is in good agreement with previous determinations based on optical and H\,I 21\,cm measurements \citep{springob05,radovich96}. The model velocity field is shown in Fig.\,\ref{figmodel}. 

\subsection{The gas kinematics}\label{gasdiscuss}

\subsubsection{The Broader Component}\label{broader}

To elucidate the origin of the broader component gas, we will compare our results with an analysis of the ionised gas in the inner 1\arcsec\ of NGC\,1667 performed by \citet{fischer13} based on HST long-slit spectra obtained with the Space Telescope Imaging Spectrograph (STIS). \citet{fischer13} measured the H$\alpha$ line and they argued that two Gaussians were needed to adequately fit the profile. They measured the velocities and velocity dispersions of each component and they argued that both are due to a biconical outflow with a position angle of the bicone of 55\ensuremath{^\circ}, an opening angle of 58\ensuremath{^\circ} and an inclination in relation to the line of sight of 18\ensuremath{^\circ}. One of the components is narrow, with velocity dispersions of $\approx$\,100\,km\,s$^{-1}$, while the other is broader, with velocity dispersions of the order of 250\,km\,s$^{-1}$. They observed redshifted velocities SW of the nucleus and a mix of redshifted and blueshifted velocities to the NE. These results are mostly consistent with our observations and, considering that the outflowing velocities of the H$\alpha$ emitting gas observed by \citet{fischer13} drop to zero at a radius of 0\farcs5 and that our observations were obtained with 0\farcs8 seeing, we argue that the differences from our results are likely due to our limited spatial resolution. Thus, we conclude that the broader component observed by us is consistent with the model proposed by \citet{fischer13} in which it originates in a biconical outflow, which is unresolved in our data.

\subsubsection{The Narrower Component}
\begin{figure*}
\includegraphics[scale=1.2]{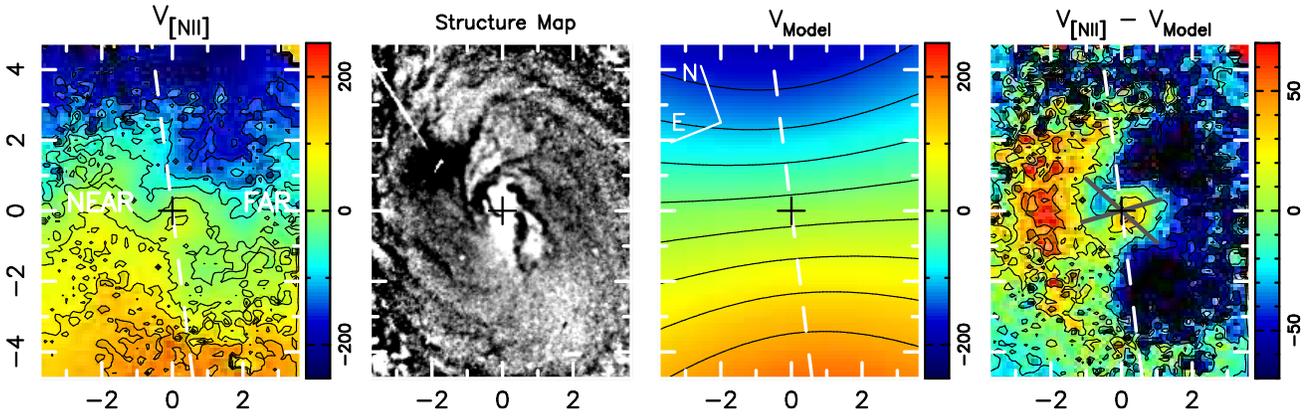}
\caption[Gaseous Kinematics]{From left to right: gaseous centroid velocity (km\,s$^{-1}$), structure map, modelled velocity field and residual between gaseous centroid velocity and modelled velocity field (km\,s$^{-1}$). The dashed white line marks the position angle of the major axis of the galaxy and the straight grey lines delimit a biconical outflow along a position angle of 55\ensuremath{^\circ} and with an opening angle of 58\ensuremath{^\circ}. The thin line in the top left corner of the structure map is an artifact.}
\label{figmodel1}
\end{figure*}

As evidenced by the centroid velocity maps shown in Fig.\,\ref{fignarrow}, the narrower component velocity field within the inner 2\arcsec\ ($\approx$\,600\,pc) is dominated by non-circular motions. To isolate these non-circular motions we need to subtract the rotation component. The rotational model we obtained for the stellar velocity field in Sec.\,\ref{discuss-stars} is not an adequate description of the gaseous velocity field, as the stellar component usually rotates more slowly than the gas \citep{vanderKruit86,bottema87,noordermeer08}. To acccount for this difference, we fitted the model described in Sect.\,\ref{discuss-stars} to the narrower component velocity field, with the velocity curve amplitude $A$ as the only free parameter in the fit. The other parameters were fixed as equal to those adopted or derived from the fit of the stellar velocity field.

The resulting amplitude of the velocity curve $A$ is $474\,\pm$7\,km\,s$^{-1}$. The model velocity field is shown in Fig.\,\ref{figmodel1}. To account for the possibility of the gas disk having a different orientation than the stellar disk, we performed a second fit, this time with the position angle of the line of nodes as an additional free parameter. This fit resulted in position angle of the line of nodes of 150\ensuremath{^\circ}, a difference of only 15\ensuremath{^\circ} compared to the large scale major axis. Additionally, the structures observed in the residual map are present and similar for both models. We therefore conclude that there is no indication in our data that the nuclear gas disk has a different kinematic position angle from that of the large scale photometric major axis. We thus adopt an angle of 165\ensuremath{^\circ} as the kinematic major axis, as the 15\ensuremath{^\circ} difference between the best fitting position angle and the large scale photometric major axis can result from a bias in the fit due to the strong non-circular motions in the gaseous velocity field.

A bar has been previously observed in the nuclear region of NGC\,1667 (\citealt{laine02}, it is visible in the continuum image in Fig\,\ref{fig1} roughly along the N-S direction), and the residual velocities observed in the inner 1\arcsec, which are of the order of 70\,km\,s$^{-1}$, are consistent with what is expected from non-circular motions in a barred potential. However, as previously discussed in Sec.\,\ref{broader}, HST long-slit spectra revealed a biconical outflow along the same orientation in the inner 0\farcs5 of NGC\,1667, thus these residuals velocities can also be due to outflowing gas. To better distinguish between outflowing, rotating, and other gas kinematics, we delineate the posited biconical outflow of \citet{fischer13} (PA\ensuremath{^\circ} 55, opening angle 58\ensuremath{^\circ}) in the residual map of Fig.\,\ref{figmodel1}. This makes it clearer that the residuals seen in the inner 1\arcsec\ - redshifted residuals observed SW on the far side of the galaxy and blueshifted residuals observed NE of the nucleus on the near side - are also consistent with this biconical outflow. Considering \citet{fischer13} also observed both a broader and a narrower component in their data and taking into account that the small spatial scale and the spatial distribution of the velocity residuals are consistent with the posited biconical outflow of \citet{fischer13}, we argue the velocity residuals in the inner 1\arcsec\ are due to a biconical outflow.

At radii between 1\arcsec and 2\arcsec the residual map reveals a two spiral pattern, a redshifted spiral on the near side of the galaxy and a blueshifted spiral on the far side. A comparison with the structure map in Fig.\,\ref{figmodel1} shows that several dust lanes are observed cospatially to the spiral pattern in the residual map. Assuming the gas is in the plane of the disk, blueshifted residuals on the far side of the galaxy and redshifted residuals on the near side imply we are observing radial inflow at the nuclear spirals. Residuals cospatial to the nuclear bar can also be due to non-circular orbits in the bar. However as the bar seems weak (the dust lanes do not trace it), it is not clear that non-circular orbits in the bar contribute significantly to the observed residuals. The velocity dispersion increase observed along the spiral arms (Fig.\,\ref{fignarrow}) supports the presence of shocks in these regions, which cause a loss of angular momentum and lead to gas flowing in.

To illustrate the scenario we proposed for the narrower component, we built a toy model of a rotating disk, two spiral arms and a biconical outflow using the software \textsc{Shape} \citep{shape}. As the overall contribution of non-circular motions along the bar to the gaseous velocity field is unclear, we do not include the bar in our model. We derived the line-of-sight velocity field from a combination of:
\begin{itemize}
 \item disk rotation;
 \item rotation and radial inflow at the location of the spiral arms;
 \item a biconical outflow;
\end{itemize}

\begin{figure*}
\includegraphics[scale=1.2]{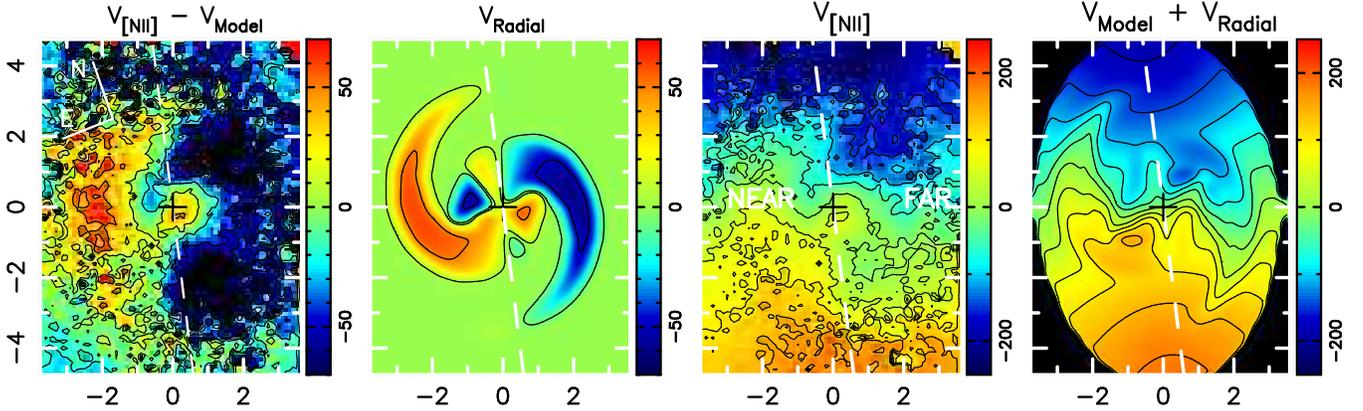}
\caption[Gaseous Kinematics]{From left to right: Residual velocities after subtraction of a rotational model from the observed gaseous velocity field, radial component of the toy model velocity field (km\,s$^{-1}$), observed gaseous centroid velocity field (km\,s$^{-1}$) and toy model velocity field (radial plus rotational components). The dashed white line marks the position angle of the major axis.}
\label{figshape}
\end{figure*}

We then convolved this velocity field with a Gaussian with a FWHM equal to that of the seeing disk. A comparison between the observed velocity field and residuals and the toy model velocity field and radial velocity is shown in Fig.\,\ref{figshape}. It is noteworthy that with such simple assumptions the toy model provides a good reproduction of the observations. 

\subsubsection{Estimating the mass outflow rate}

The mass flowing through a cross section of radius $r$ can be obtained from:

\begin{equation}
\dot{M}_{out}\,=\,N_{e}\,v\,\pi\,r^{2}\,m_{p}\,f
\end{equation}

where $N_{e}$ is the electron density, $v$ is the velocity of the inflowing gas , $m_{p}$ is the mass of the proton, $r$ is the cross section radius and $f$ is the filling factor. The filling factor can be estimated from equation (2).
Substituting equation (2) into equation (4), we have for a mass flow through a cone of height $h$ and radius $r$:

\begin{equation}
\dot{M}_{out}\,=\frac{3\,m_{p}\,v\,L_{H\alpha}}{j_{H\alpha}(T)\,N_{e}\,h}
\end{equation}

where $L_{H\alpha}$ is the H$\alpha$ luminosity, $J_{H\alpha}(T)$\,=\,3.534$\,\times\,10^{-25}$\,erg\,cm$^{3}$\,s$^{-1}$ for a temperature of 10000K \citep{osterbrock06}, and $h$ is the height of the bicone (equivalent to the distance from the nucleus). As the broader component is unresolved in our data, we will base our estimates on the narrower component measurements.
We will adopt the density, outflowing velocity and $L_{H\alpha}$ values obtained from the redshifted region extending from the nucleus to 1\arcsec\ (296\,pc) to the SW. The average electron density in this region is 670\,cm$^{-3}$ (from the [S\,II] ratio). The total H$\alpha$ flux in this region is $3.6\,\times\,$10$^{-14}$\,erg\,cm$^{-2}$\,s$^{-1}$. Adopting a distance of 61.0\,Mpc, we obtain a total luminosity for the outflowing gas of $L_{H\alpha}$\,=\,$1.6\,\times\,$10$^{40}$\,erg\,s$^{-1}$.  Adopting a distance from the nucleus (height of the cone) of 1\arcsec, we obtain an average projected outflowing velocity of $\approx$\,41\,km\,s$^{-1}$. Correcting this velocity by an inclination of 18\ensuremath{^\circ} \citep{fischer13}, we obtain a velocity of 132\,km\,s$^{-1}$. We thus obtain an outflow mass rate of $\dot{M}_{out}$\,$\approx$\,0.16\,M$_{\odot}$\,yr$^{-1}$ for the narrower component. This value is in good agreement with those from previous studies which estimated mass outflow rate in nearby AGNs (z\,$<0.1$) ranging from $\approx$\,0.01\,M$_{\odot}$\,yr$^{-1}$ to $\approx$10\,M$_{\odot}$\,yr$^{-1}$ \citep{rogemar11a,rogemar11b,sanchez11,allan14a,schonell14,allan16,karouzos16}.

\subsubsection{Estimating the mass inflow rate}
\label{inflow}

\begin{figure*}
  \centering
  \includegraphics[scale=0.5]{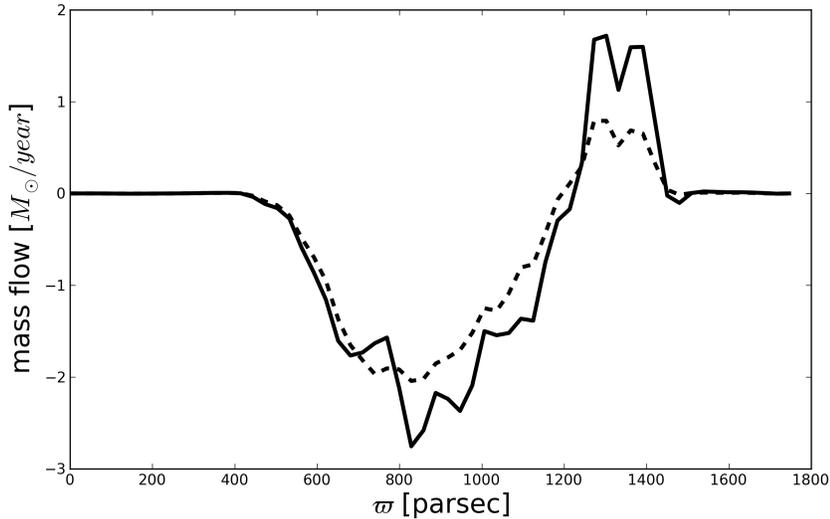}
  \caption{Mass flow rate $\dot{M}$ obtained from the residual velocity field (after subtracting the rotation model -- solid curve) and from the observed velocity field (without subtracting the rotation model -- dashed curve).}
  \label{figure12}
\end{figure*}

In this section we describe our method to estimate the mass flow rate towards the nucleus: the integration of the ionised gas mass flow rate through concentric rings in the plane of the galaxy (around the whole perimeter). In order to do this, we consider that the observed line-of-sight velocity, $v_{LOS}$ is the result of the projection of three velocity components in a cylindrical coordinate system at the galaxy. The cylindrical  coordinates are: $\varpi$ -- the radial coordinate; $\varphi$ -- the azimuthal angle;  $z$ -- the coordinate perpendicular to the plane. The galaxy inclination relative to the plane of the sky is $i$. We assume that gas vertical motions in the disc are negligible, i.e. $v_z= 0$, and then we consider two possibilities, as follows. 

We first consider that the azimuthal velocity component $v_\varphi$ can be approximated by the rotation model we fitted to the velocity field, and subtract it from $v_{\rm LOS}$ in order to isolate the radial velocity component. As the de-projection of the resulting radial component into the plane of the galaxy is not well determined along the galaxy line of nodes due to divisions by zero, we have masked out from the velocity field a region of extent 0\farcs3 to each side of the line of nodes in the calculations.  

The gas mass flow rate is given by:
\begin{equation}
\label{eq:mass_flow_v.A}
\dot{M} = \rho f \  \mathbf{v}\cdot\mathbf{A},
\end{equation}
where $\rho$ is the ionised gas mass density, $f$ is the filling factor (determined via Eq.\,2), $\mathbf{v}$ is the radial velocity vector and $\mathbf{A}$ is the area vector through which the gas flows. Since we are interested in the radial flow, the area we are interested in is perpendicular to the radial direction, and thus $\mathbf{v}\cdot\mathbf{A} = v_\varpi\, A$, the product between the area crossed by the flow and the radial velocity component.

The filling factor $f$ is estimated from:

\begin{equation}
L_{H\alpha}\,\sim\,f\,N_{e}^{2}\,j_{H\alpha}(T)\,V
 \end{equation}
where $j_{H\alpha}(T)$\,=\,3.534$\,\times\,10^{-25}$\,erg\,cm$^{-3}$\,s$^{-1}$ \citep{osterbrock06} and $L_{H\alpha}$ is the H$\alpha$ luminosity emitted by a volume $V$.  We consider a volume $V$\,=\,$A\,dx$, of a thick ring sector with width $dx$ and cross-section area A, the same as above. Replacing $f$ obtained via Eq.\,2 into Eq.\,1,  we finally have:
\begin{equation}
\dot{M}(\varpi) = \frac{m_p \, v_{\varpi} \, L_{\rm H\alpha}}{j_{H\alpha}(T)\, N_e \,dx}.
\label{eq:dotm}
\end{equation}

Integrating Eq.\,~\ref{eq:dotm} all around the ring of radius $\varpi$, and taking into account the radial and azimuthal dependence of $v_{\varpi}$, $L_{\rm H\alpha}$ and $N_e$, we obtain the gas mass flow rate $\dot{M}$ as a function of the radius $\varpi$. It is worth noting that the mass flow rate as calculated above does not depend on $A$ and $dx$, as these quantities cancel out in Eq.\,~\ref{eq:dotm}, when we use $L_{\rm H\alpha}$ as the gas luminosity of a volume $V$\,=\,$A\,dx$.

We have evaluated $\dot{M}(\varpi)$ through concentric Gaussian rings (rings whose radial profile is a normalised Gaussian) from the galaxy centre up to the maximum radius of 1.8\,kpc (in the galaxy plane). The result is shown in Fig.\ref{figure12}. The negative values mean that there is inflow from $\approx$\,1.2\,kpc down to 400\,pc from the nucleus. With the mass inflow rate being the largest between 1\,kpc and 800\,pc and then decreasing until it reaches $\approx$\,0 at 400 pc from the nucleus. It is worth noting, however, that in the inner 400\,pc (1\farcs3) the residuals are mainly close to the major axis, as can be seen in the residual map in Fig.\,\ref{figmodel}, and as we have masked out from the velocity field a region of extent 0\farcs3 to each side of the line of nodes, so a significant part of the residuals are not considered in the mass flow rate estimate. Additionally, it is possible that the gas inflow continues to smaller radii, but in the case of much stronger emission from the outflowing gas compared to inflowing gas, the inflowing gas would be undetectable by our observations.  

A second approach is to calculate the inflow rate without subtracting the rotational component. One might indeed question if subtracting the rotational component is necessary, as its contribution to $v_{\rm LOS}$ may cancel out in the integration of the mass flow rate around the ring if the gas density and filling factor have  cylindrical symmetry, as by definition, the rotational component has cylindrical symmetry.

We now compare the estimated inflow rate of ionised gas to the mass accretion rate necessary to produce the luminosity of the Seyfert nucleus of NGC\,1667, calculated as follows:
\[
\dot{m}\,=\,\frac{L_{bol}}{c^{2}\eta} 
\]
where $\eta$ is the efficiency of conversion of the rest mass energy of the accreted material into radiation. For geometrically thin and optically thick accretion disk , the case of Seyfert galaxies, $\eta$\,$\approx$\,$0.1$ \citep{frank02}. The nuclear luminosity can be estimated from the X-ray luminosity of $L_{X}$\,=\,2.0\,$\times\,$10$^{42}$\,erg\,s$^{-1}$ \citep{panessa06}, using the approximation that the bolometric luminosity is $L_{B}$\,$\approx$\,10$L_{X}$. We use these values to derive an accretion rate of $\dot{m}$\,=\,3.5$\,\times\,$10$^{-3}$\,M$_{\odot}$\,yr$^{-1}$.

Comparing the accretion rate $\dot{m}$ with the mass inflow rate of ionised gas obtained in our analysis above, we find that in the inner 600\,pc the inflow rate is 2 orders of magnitude larger than the accretion rate, reaching a maximum mass flow rate of $\approx$\,2.8\,M$_{\odot}$\,year$^{-1}$ at 800\,pc from the nucleus. We point out that this inflow rate corresponds only to ionised gas, which is probably only a fraction of a more massive inflow, as ALMA (Atacama Large Millimeter Array) observations of NGC\,1667 shows that the molecular gas is also inflowing along the nuclear spirals (Slater et al., in preparation).  

The orders of magnitude difference between the accretion rate and the mass inflow rate raises the question of which is the fate of inflowing the gas. Considering our result that the gas inflow rate reachs a maximum at $\approx$\,800\,pc and then decreases at smaller radii, we conclude that only a small fraction is likely to reach the nucleus. Most of it will probably accumulate in the inner few hundred parsecs. A comparison with the nuclear region of NGC\,1433 \citep{combes13}, which shows remarkably similar structures to those in the nuclear region of NGC\,1667 (a multi spiral pattern and a nuclear bar inside a nuclear ring) supports this scenario. In NGC\,1433 the molecular gas does not follow the nuclear bar; instead it flows inward along the nuclear spiral arms and partly accumulates in a ring-like structure at a radius of $\approx$\,200 pc \citep{combes13}. \citet{combes13} pointed out that such gas inflows are expected at some epochs of self-consistent N-body+hydro simulations, when the gas enters an inflowing phase inside two inner Lindblad resonances.

There is evidence that gas has been accumulated in the inner few hundred parsecs of nearby Seyfert galaxies in the last 10$^{6}$--10$^{8}$\,yrs. Regions of low stellar velocity dispersion have been observed in the inner $\approx$\,200\,pc of nearby Seyferts \citep{emsellem08,comeron08} and young to intermediate (10$^{6}$--10$^{8}$\,yrs old) stellar population have been observed associated to these regions \citep{riffel10,riffel11,thaisa12,hicks13}.  These low velocity dispersion regions have been interpreted as due to a stellar component in the vicinity of the nucleus of Seyfert galaxies that is dynamically colder in comparison to the stars that dominate the detected stellar light in quiescent galaxies \citep{hicks13}. Consequently, a reservoir of gas must have build up in the inner few hundred parsecs of Seyfert galaxies from which new stars could form. We might be observing the build up of this reservoir in NGC\,1667.

\section{Summary and Conclusions}\label{Conclusion}

We have measured the stellar and gaseous kinematics of the inner 2\,$\times$\,3\,kpc$^2$ of the Seyfert\,2 galaxy NGC\,1667, from optical spectra obtained with the GMOS integral field spectrograph on the Gemini South telescope at a spatial resolution of $\approx$\,240\,pc. The main results of this paper are:

\begin{itemize}

\item The stellar velocity field shows velocity dispersions of up to 400\,km\,s$^{-1}$ and circular rotation consistent with an orientation for the line of nodes of $\approx$\,165$^\circ$;
 
\item Extended gas emission is observed over the whole FOV, with the line profiles being well fitted by Gaussian curves; 

\item In the inner $\approx$\,2\arcsec, two gaseous kinematical components are needed: a narrower component, that is present over the entire FOV, and a broader component;

\item We interpret the broader component as due to an unresolved biconical outflow;

\item The narrower component velocity field shows strong non-circular motions. The subtraction of a rotational model shows a two spiral pattern with residual velocities of the order of 60\,km\,s$^{-1}$ between 1\arcsec and 3\arcsec\ from the nucleus. Spiral dust lanes are observed cospatially to the spiral pattern in the residual map;

\item We observe residual redshifts in the spiral arm in the near side of the galaxy and blueshifts in the spiral arm in the far side. We interpret these residuals as radial inflows;

\item We argue that shocks in the gas in the nuclear spirals cause loss of angular momentum and leads to the gas inflow. The presence of shocks is supported by increased velocity dispersion observed in the arms;

\item In the inner 1\arcsec\ redshifted residuals are observed NE of the nucleus in the far side of the galaxy and blueshifted residuals are observed SW, on the near side. We interpret these residuals as due to outflowing gas;

\item We show that the observed velocity field can be reproduced by a toy model including disk rotation, radial inflow associated to spiral arms and a biconical outflow;

\item We calculate the net gas mass flow rate across a series of concentric rings around the nucleus, obtaining a maximum mass flow rate of $\approx$\,2.8\,M$_{\odot}$\,year$^{-1}$ at 800\,pc from the nucleus;

\item Considering the mass flow rate decreases for radii smaller than 800\,pc, it seems that most of the inflowing gas will not reach the nucleus to fuel the AGN. Instead, it will probably accumalate in the few inner 100 parsecs and fuel the formation of new stars;

\item We calculate the mass outflow rate for the narrow component gas, obtaining $\dot{M}_{out}$\,$\approx$\,0.16\,M$_{\odot}$\,yr$^{-1}$. 

\end{itemize}

\section*{ACKNOWLEDGMENTS}

We are grateful to the anonymous referee for useful and interesting comments and suggestions which have improved this paper. This work is based on observations obtained at the Gemini Observatory, which is operated by the Association of Universities for Research in Astronomy, Inc., under a cooperative agreement with the NSF on behalf of the Gemini partnership: the National Science Foundation (United States), the Science and Technology Facilities Council (United Kingdom), the National Research Council (Canada), CONICYT (Chile), the Australian Research Council (Australia), Minist\'erio da Ci\^encia e Tecnologia (Brazil) and south-eastCYT (Argentina). This work has been partially supported by the Brazilian institution CNPq.

\bibliographystyle{mn2e.bst}
\bibliography{ngc1667.bib}

\label{lastpage}
\end{document}